\documentclass[useAMS,usenatbib]{mn2e}
\usepackage{graphicx}
\usepackage{epsfig}
\newcommand{\msol}{\mbox{M\hbox{$\odot$}}}

\title{Probing the magnetic fields of
  massive star forming regions with methanol maser polarisation}
\author[Dodson, Moriarty]{R. Dodson$^{1} $\thanks{E-mail: richard.dodson@icrar.org},  C. D. Moriarty$^{1}$\\
$^{1}$ International Centre for Radio Astronomy Research, Perth, Australia\\
}
\begin{document}

\date{Accepted . Received .}

%\pagerange{\pageref{firstpage}--\pageref{lastpage}} \pubyear{2011}

\maketitle

%\label{firstpage}

\begin{abstract}
Methanol masers can provide valuable insight into the processes
involved in high-mass star formation, however the local environment in
which they form is still unclear. 
Four primary, yet conflicting, models have emerged to explain the
commonly observed methanol maser structures at 6.67 GHz. These suggest
that masers trace accretion disks, outflows, shock fronts or disks
dominated by infall/outflows. One proposed means of testing these models is
through mapping the local magnetic field structures around maser
sources, which were predicted to lie parallel to shock and outflows
and perpendicular to accretion disks.
To follow up this suggestion we have determined magnetic field
directions from full polarisation observations of 10 6.67-GHz sources.
We find morphology that is parallel to the source structure,
indicative of shocks or outflows, in five sources and perpendicular
morphology indicative of disks in three.
These results do not support any of the expected models and
the diverse morphologies observed indicate that the masers could be
emitting from different evolutionary stages or environments, or from a
common local environment with complex associated magnetic fields. 
To resolve this conflict we suggest a new approach that will search
the simulations of massive star formation, which are just becoming
available, for suitable sites for maser emission.
\end{abstract}


\onecolumn
\section{Introduction}
The processes involved in high-mass star formation remain the subject
of significant ongoing debate. In particular, it is unclear whether
the accretion model currently accepted for lower mass stars ($<$8
M$_{\odot}$) is still applicable at higher masses, or whether an
alternate mechanism is required.  For massive stars the accretion-disc
model becomes more complex, as radiation pressure produced by the
protostar becomes theoretically large enough to halt the gravitational
inflow of matter before the final mass of the star is achieved
\citep{Zinnecker07}.  The current models of massive star 
formation (MSF) are difficult to verify due to the relative scarcity of
high-mass star forming regions and the fact that many are obscured by
dense surrounding molecular clouds, preventing direct observation.  One
means of probing these sources is through studies of their radio
emission emitted by the molecular clouds surrounding the central
site. 6.67-GHz methanol masers are an abundant and intense maser
species known to be associated with sites of high-mass star
formation \citep{minier05}. This makes them ideal probes of the processes at
work in their environment.

Current methanol maser studies have focused on the conflicting
theories regarding local maser environment. By determining whether
methanol masers can be associated with accretion disks, or potentially
other star formation mechanisms, they can illuminate the high-mass
star formation debate. Studies by \cite{Norris93,Norris98} suggested
that the observed linear spatial distributions and velocity gradients
of methanol maser features across a source were tracers of rotating
accretion disks, which was considered strong evidence that the
accretion disks underpinned high-mass star formation.  This theory has
been further supported through more recent work by \cite{Pestallozi04}
(on NGC7538) 
and \cite{Mosca11} (on IRAS\,20126+4104). 
\cite{Surcis11} has recently advanced a variation of the model for
NGC7538, which has the masers on the interface between infall and a
torus.
However, a wide survey by \cite{Walsh98} found
only 36 out of 97 methanol maser sources to be in linear structures,
with the remainder being inconsistent with the disk
model. Furthermore, the estimates for the central masses that were
derived through a radial analysis of the linear methanol masers
yielded extremely large and unlikely stellar masses (M $>$
100M$_{\odot}$) in some cases. Walsh postulated an alternate model to account
for the diverse properties of the methanol masers observed in their
survey, in which masers are created through the propagation of shock
fronts through molecular clouds.
\cite{Dodson04} further suggested that the masers might be occurring as
external shocks propagating through a rotating molecular cloud, which
would account for the observed linear velocity gradients. However
\cite{vanderWalt07} felt that, after considering typical rotation
rates of molecular clouds, the shock front model could not account for
the small velocity range observed across maser observations. Shock
fronts would be expected to propagate at high speeds, which are
inconsistent with the observed low velocities across methanol
masers. This overlooks the requirement that, to
allow for a long line-of-sight coherent velocity path, the shocks must
be close to the plane of the sky, which would reduce the observed
velocity range.
An alternative model that has been suggested is that methanol
masers are located within bipolar outflows. The maser emission is in
an environment in which water masers are known to be produced
\citep{Bartkiewicz11}, and these can be found to trace bipolar
outflows \citep{Moscadelli05}. Support for the model was found by 
\cite{deBuizer03}, who observed H$_{2}$ emission, and in
\cite{deBuizer09} observed SiO using the JCMT (SiO 6$\rightarrow$5)
and ATCA (SiO 2$\rightarrow$1). Both the SiO and H$_2$ tracers are
typically associated with outflows, and tended to be aligned parallel
to the methanol maser distributions. However \cite{Bartkiewicz09}
showed that attempts to apply the model by analysing the observed
velocities of maser spots led to a cone vertex that would be offset
from the central stellar mass. Additionally, although this model
accounts for linearly distributed maser spots, the extent of the
linear distributions suggest that the outflows associated with them
would need to be highly collimated, requiring high velocities,
inconsistent with those observed.
In their studies \cite{Bartkiewicz09} find that a significant
fraction of the sources they study form rings with only a few of the
sources being linear, which they credit to the improved sensitivity of
their observations. Their best explanation of the sources is that they
arise from masers on expanding and rotating disks. %, but their model 
%would require proper motion monitoring to be confirmed. 
%

Several programs of phase referenced VLBI observations of methanol
masers are underway, e.g. \cite{Sanna10a,Sanna10b}, which will
determine the kinematics of these sources. Such observations will
provide a powerful diagnostic, but require a long observational
timespan. 
Broad-scale magnetic field morphology should provide an immediate test of the
models of methanol maser environments. In the accretion model,
magnetic fields follow the gravitational collapse of molecular clouds
around a protostar and are expected to thread through the resulting
accretion disk in an hourglass shape \citep{Lizano09}. Thus, if linear
maser features traced an accretion disk, magnetic field vectors would
be expected to appear perpendicular to the major axis. Conversely,
linear distributions in both outflows and shock fronts would be
expected to lie parallel to the magnetic field vectors.  
Methanol is a non-paramagnetic molecule, however the application of a strong
magnetic field during the pumping process can lead to small fractional
linear or circular polarisation. As a result, the polarisation of
observed maser radiation can be directly attributed to the influence
of a strong and ordered magnetic field across the maser site. Previous
methanol maser studies, (e.g. \cite{Vlemmings06},
\cite{Dodson08}, \cite{Surcis09}, \cite{Vlemmings10}) have found that
the magnetic field vectors show an alignment across the source that
was not consistent with the disk model. However the sample size of
current polarisation analysis has been too small to observe the extent
of this discrepancy. Focusing on the polarisation properties of a
sample of 10 observed masers at 6.67 GHz, we have obtained estimates
of the broad magnetic field morphology at the sites housing maser
activity.
%Our deduced morphologies varied across our source sample,
%supporting the argument that methanol masers do not provide consistent
%tracers of accretion disks in high-mass star forming regions.

\section[Observations]{Observations}
The data used in this analysis came from four separate observing
sessions: 22 September 1999, 25 September 1999, 20-21 November 2002
and 27 December 2008. 
%The earlier data was taken from the archive and a further 
%session of Directors Time was granted to make confirming observations.
All data was taken with the Australia Telescope
Compact Array (ATCA) using a 4 MHz bandwidth with 1024 frequency
channels, giving a velocity resolution of 0.2 kms$^{-1}$.  They were
performed with dual linear antenna feeds to allow for full
polarisation analysis, with an initial instrumental polarisation of
approximately 1\%. %\cite{atcaguide}. {\bf is this
%  correct?}  
The ATCA configurations were 6A and 6C, which typically gives a
resolution of 1 to 4 arcseconds depending on the declination.
The calibration of data was performed using the \emph{Miriad} software
package.  All pointings had 20 minutes or more of on-source
observation per session, and the primary (and bandpass) calibrations
were obtained from 30 minutes observations of PKS~1934-638. To achieve
good polarisation calibration, high quality calibration solutions are
needed for each channel, thus the unusually long bandpass scan.
The primary and secondary sources were calibrated as-per the
methods outlined in the \emph{Miriad} manual \citep{miriad}, keeping
the leakage solutions from the primaries to provide the most
consistent results. The secondary calibrators in the 2008 dataset
provided poor gain results, hence the sources from this set used both
the gains and leakage solutions from the primary calibrator
1934-638. Further corrections to the modelled antenna gains were made
by performing phase self-calibration on the maser sources in each set
across the strongest channel. For sources G291.27-0.70 and
G345.01+1.79, weak continuum radiation was visible in the spectra
after self-calibration, which was removed using the subtraction task
{\sc UVLIN}. The sources were then imaged with 0.3 kms$^{-1}$
channels and deconvolved, combining the data for the two 1999
observations, as well as combining the 1999 and 2002 observations of
sources G339.88-1.26 and NGC6334F. 
Velocity channels were fitted with simple models to provide component
positions.
The deconvolved Stokes I, Q and U spectra were exported for
polarisation analysis in \emph{Mathematica}. Plots of the fractional
linear polarisation and polarisation angle, with associated errors,
were generated, clipping points with low I intensity. A full
description of the methods and the data reduction can be found in
\cite{Moriarty09}\footnote{http://ftp.physics.uwa.edu.au/pub/Honours/2009/Theses/Moriarty}. All
the image cubes had a sensitivity within a factor of 3 of 0.3\,Jy per
channel.

\section[Results]{Results}

Before commencing the discussion of the results it is important to
consider the relationship of the linear polarisation angle $\chi$ to
the magnetic field direction. The dominant maser emission can be
either parallel or perpendicular to the magnetic field lines,
depending on the angle between the magnetic field and line of
sight. If it is greater (less) than $\sim$55$^{\circ}$ the vectors are
perpendicular (parallel) to the magnetic field
\citep{Goldreich1973}.  When the masers are highly
saturated there are further complications as non-linear effects alter
the relationships, such as the fixed
proportionality between circular polarization and magnetic field
strength.
%% Which are xxxx
%
One can deduce the emerging brightness temperature, T$_b\Delta\Omega$,
from the circular polarisation measurements, but our datasets do not
contain that information. Assuming a typical spot size of 3\,mas and a
flux of 100\,Jy the brightness temperature would be
3$\times$10$^{11}$\,K. Unless the beaming angle $\Delta\Omega$
approaches several degrees this implies that the emerging brightness
temperature is lower enough to ensure that the maser is unsaturated.
Furthermore highly saturated masers are expected to have significantly
higher fractional polarisation levels than we have observed and we can
therefore conclude that the emission is never sufficiently saturated to change the
emitting regime.
% 
% Brightness temperatures for 100 Jy 1mas = 3E12K
% Brightness temperatures for 100 Jy 3mas = 3E11K
% 
% 100e-26/2/k/b/b*l*l  ---  lambda=4.5e-2 beam=5e-9
%
If the masers arise in a disk seen edge on, as in the model of Norris (1993),
the field lines will be perpendicular to the disk, and therefore at
$90^\circ$ to the line of sight. The angle will remain greater than
$55^\circ$ until the axial ratio between the major and minor axes
reaches 57\%. As can be seen the Figures, none of our targets show such a large ratio.
Alternatively if the targets are due to outflows their extended
structure would lead one to expect that they and their magnetic fields
are not directed towards us. In which case, again, the magnetic field
angle will be perpendicular to the observed polarisation angle. 
Finally if the targets are ring-like structures dominated by outflow or infall
rather than rotation the field structure would be in the disk and
radial from the centre of mass. In this case observations from above
would show a polarisation vector that rotates across the
structure.  Observed edge-on the angle to the line of sight would be
$90^\circ$ at the extremes and aligned when looking towards the
centre of mass.
Therefore we have assumed that the angle is greater than $55^\circ$ in
all plots, as in previous work \citep{Dodson08,Surcis09}, and discuss the likelihood of it not being so in the text. 

The one region where this may not be correct is around components C
and E in the Western cluster of G339.88-1.26, where the polarisation
vectors rapidly change by 90$^\circ$ between the two adjacent emission
spots. This is more clearly revealed in the VLBI image presented in
\cite{Dodson08}.

The morphology of each maser region has been studied through
comparison of our estimates of the magnetic field vector directions
and the linear direction of the source. To estimate magnetic field
vectors in the maser regions from the polarisation results, we rotated
the polarisation angle ($\chi$) to account for a galactic magnetic
field correction ($\phi$) as well as the assumed 90$^{\circ}$ angle difference
between the polarisation and magnetic field angles. Local galactic
rotation measure (RM) model values for each source are from 
\cite{VanEck10} using the directions and lower-limit kinematic
distances. These were nearly always small and are listed in Table
\ref{tab:Summary-of-Source}. 
The RM models are based on a limited number of measurements
extrapolated on to the lines of sight for our sources, whereas the
masers are in dense high-mass star forming regions that could be
expected to show significant deviations from a global
model. Nevertheless the model is the best approach we have to find the
estimates for the RM, and the corrections are sufficiently small that
errors inherent in them will not make a significant contribution.
Magnetic field angles $\theta_{B}$ for each source were
calculated by subtracting the expected RM angle $\phi$ and
rotating to obtain the perpendicular vectors

\begin{center} 
\begin{equation}
  \vspace{2cm}  \theta_{B}=\chi-\phi+90
  \end{equation} 
\end{center}

To determine the relationship between the magnetic field angles and the
distribution of maser spots, $\theta_{B}$ results were mapped against
position data for individual velocity features.

\subsection{Magnetic field morphologies}
\begin{figure}
\vspace{0.3cm}
%\hspace*{-2cm}\includegraphics[width=12cm]{291spots.pdf}
\hspace*{-2cm}\includegraphics[width=12cm]{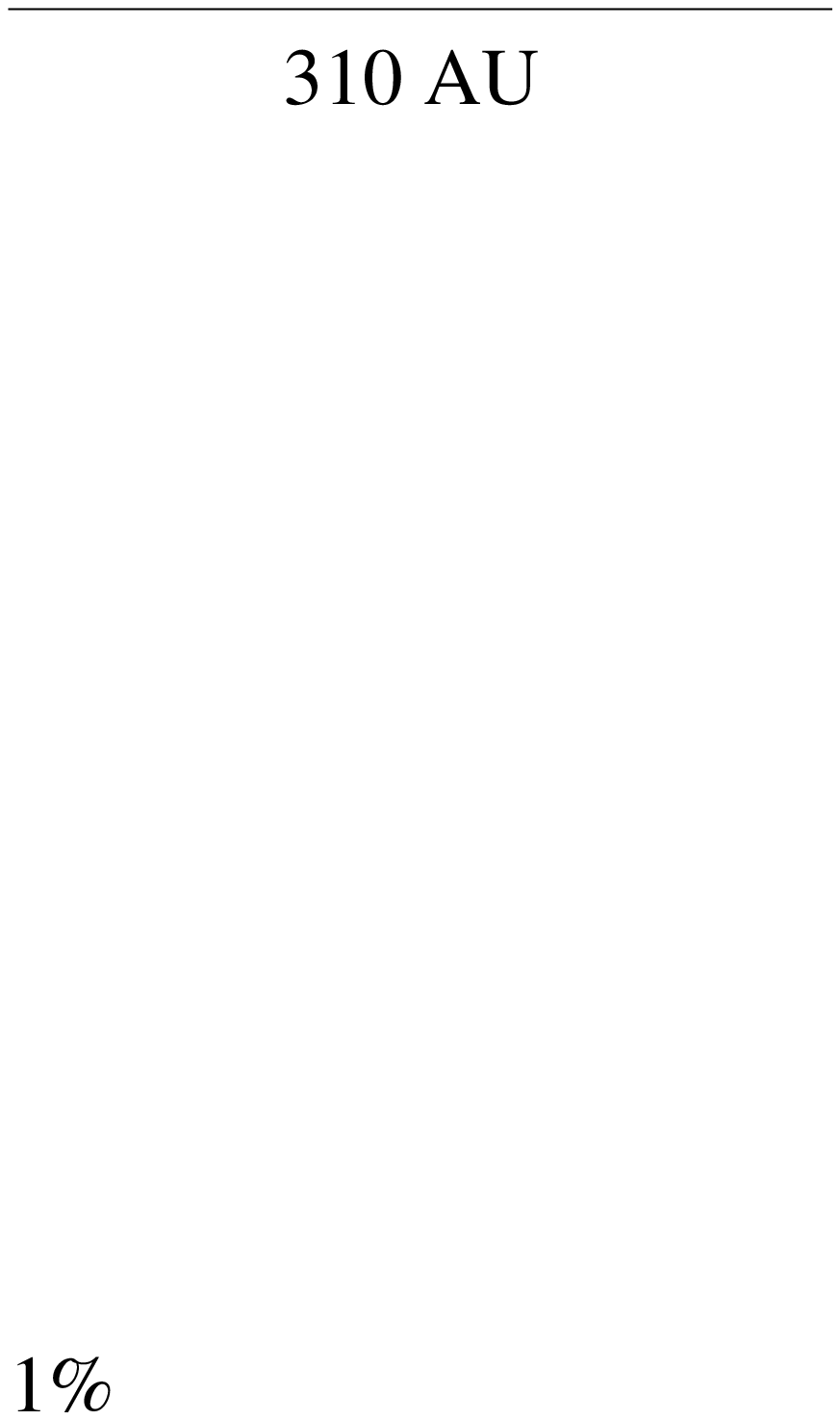}
%\vspace{-4cm}

\includegraphics[width=8cm]{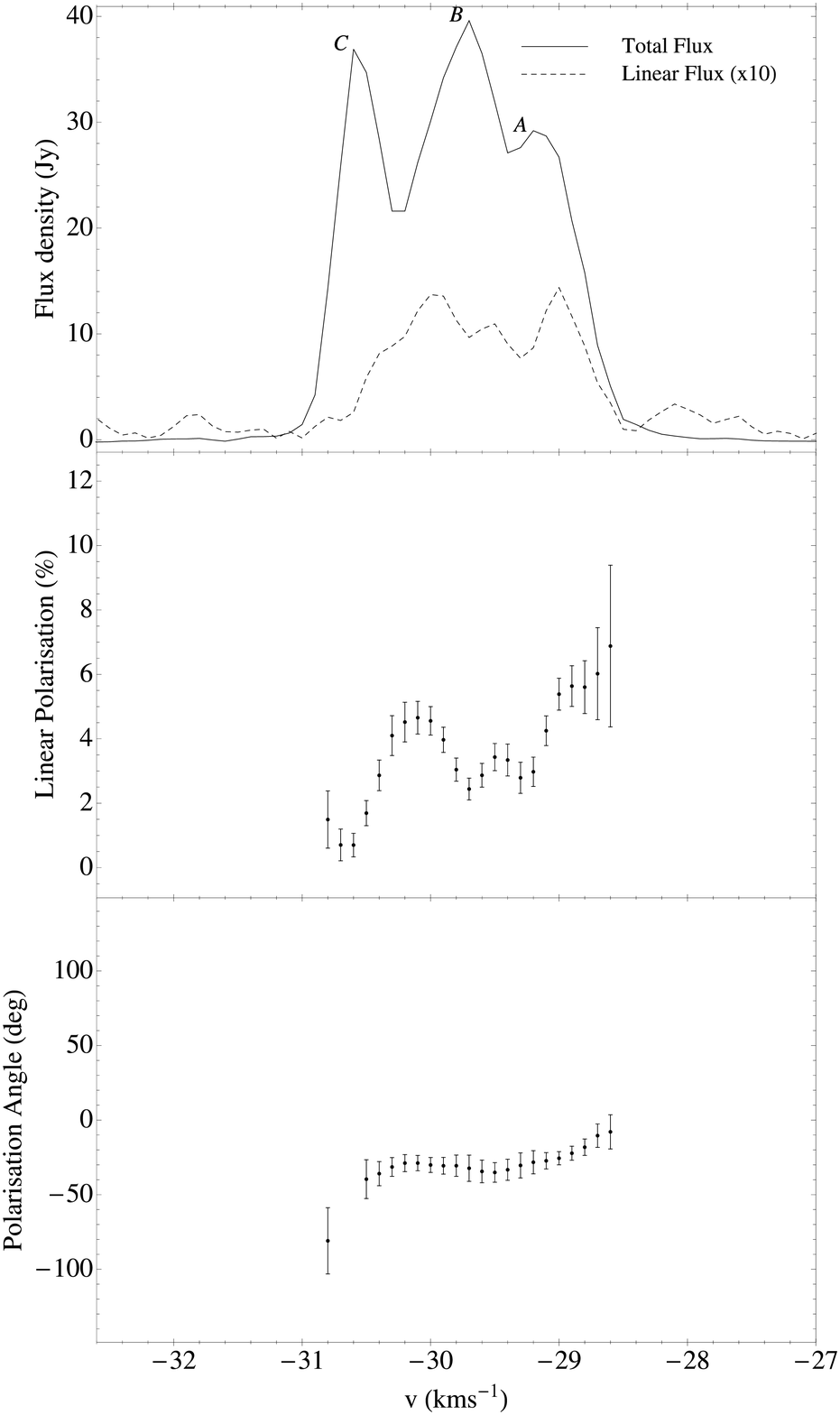}
\caption{}\label{fig:0}
\end{figure}

\begin{figure}
\includegraphics[width=8cm]{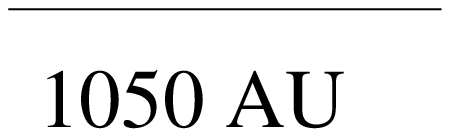}
%\end{figure}

%\begin{figure}
\includegraphics[width=8cm]{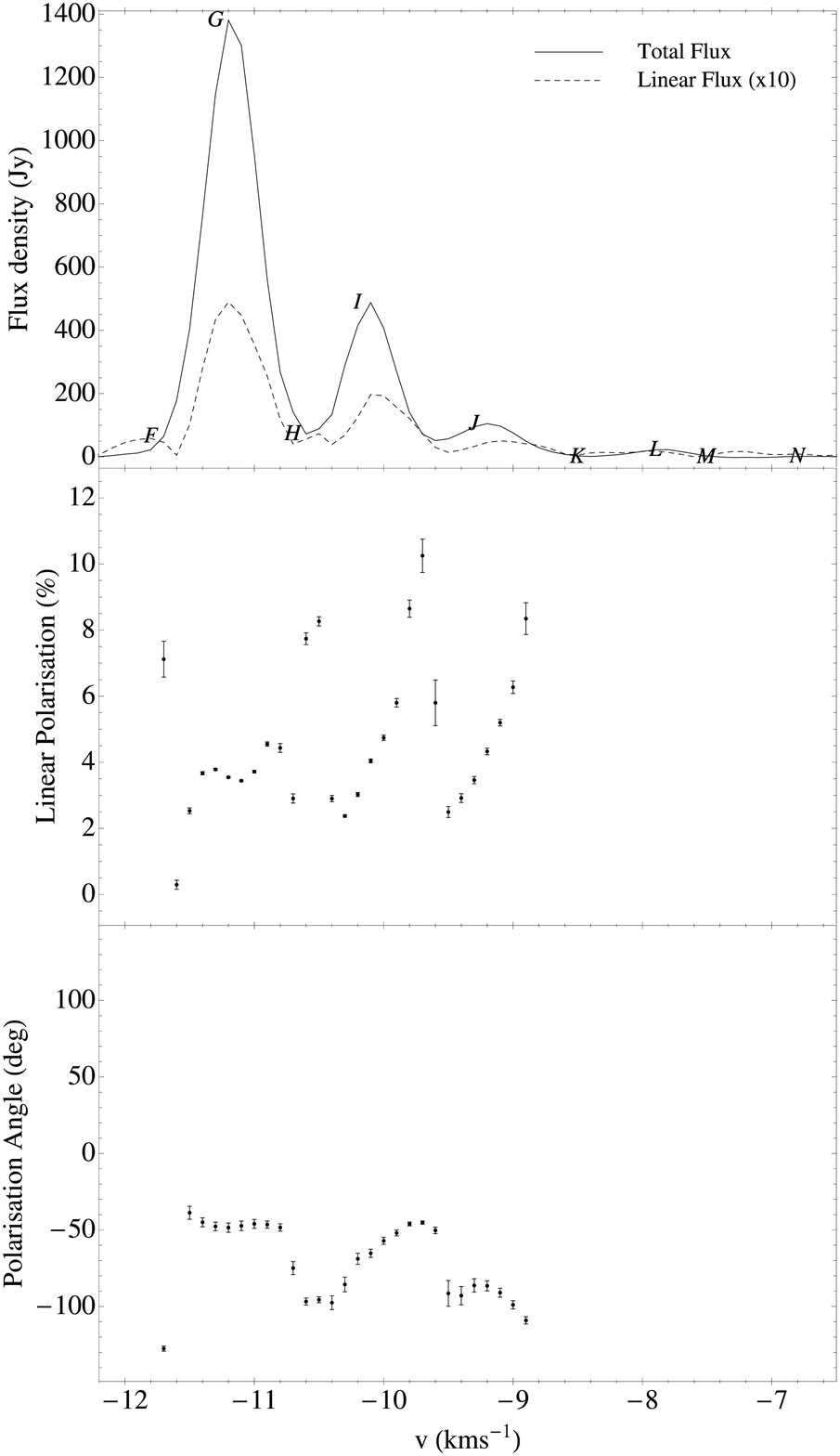}
\caption{As before, for NGC6334F (NW), relative to 17:20:53.45 -35:47:00.5}\label{fig:9}

\end{figure}

%\typeout{Re-read this!!}{\bf re-read this} 
Once values of $\theta_{B}$ had been determined for the individual
spots they were compared to the major axis angles of the maser spot
distributions ($\theta_{axis}$) and classified based on the angle
offset between the two. We divided the results between the two
classes. Angle differences closer to parallel
($|\theta_{B}$-$\theta_{axis}|<45$) were classified as parallel, and
angles closer to perpendicular
($|\theta_{B}$-$\theta_{axis}|>45$) were classed as
perpendicular. Some sources were classified as disordered based on
their morphology.
%
%Similar conclusions would be obtained if we divided into three classes of parallel, intermediate and perpendicular, each covering 30$^\circ$. 
%
Some
sources have components that fall under all classifications
(e.g. G339.88-1.26), and in these cases the dominant classification
was taken. Major axes were determined through lines of best fit,
weighted by maser flux, where the addition of the weighting constraint
provided the best approximation to a linear axis for the maser
sources.  The results are presented in Figures \ref{fig:0} through to
\ref{fig:9}.
It should be noted that this simplified classification does not
provide a complete description of the often-complex nature of magnetic
field morphologies. However, it provides a test of the primary models
for the environments that house linear distributions of methanol
masers, which have clearly associated perpendicular/parallel expected
field vectors. Looking at the overall morphology trends for the ten
sources, %for which magnetic field vectors were able to be mapped, 
three
of them showed predominantly perpendicular vectors to the maser axes,
five displayed parallel vectors and two displayed disordered
structure. A table of the associated properties for each source is
provided in Table \ref{tab:Summary-of-Source}, which lists (amongst
other parameters) in column 2 the classification based on the
individual spots, and in column 8 the offset between the average polarisation
vectors and the weighted major axes.
%
%For G339.88-1.26, because of the averaging over the
%complicated structure, the two classifications would appear 
%inconsistent.

%table 1 here

\subsection{Individual sources}
\subsection*{G291.27-0.70}
G291.27-0.70 is a relatively weak and compact maser, peaking at 39.6\,Jy
and having an extent of only a few hundred mas. However
following the method in \cite{Phillips98} we were able to fit
positions to the velocity channels with sufficient accuracy to map the
source.
The results are presented in Figure \ref{fig:0}, which shows the maser
spot positions on the sky, with the polarisation angles (where
detected) overlaid, rotated to indicate the magnetic field
direction. The 1-$\sigma$ errors define a range of angles, which are
indicated with red dashed lines. The fitted major axis is shown with a
light solid line, to indicate the relative orientation of the field direction
and the source major axis. In addition the velocity spectrum is shown
with total and polarized flux densities, the percentage linear
polarisation and the polarisation angle, where detected.
The magnetic field angles on the plane of the sky of G291.27-0.70 were found to be constant
across the source, and lie at $\theta_{B}$-$\theta_{axis}$=-51$^\circ$ (perpendicular by our
definition) to the source major axis. This we would
consider consistent with the model of methanol masers tracing an
accretion disk. However the small size and velocity dispersion leads to a
predicted enclosed mass of only 0.2 \msol .

\subsection*{G305.21+0.21}
\begin{figure}
\includegraphics[width=8cm]{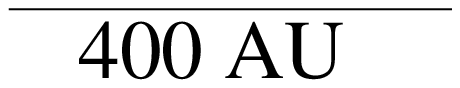}

\includegraphics[width=8cm]{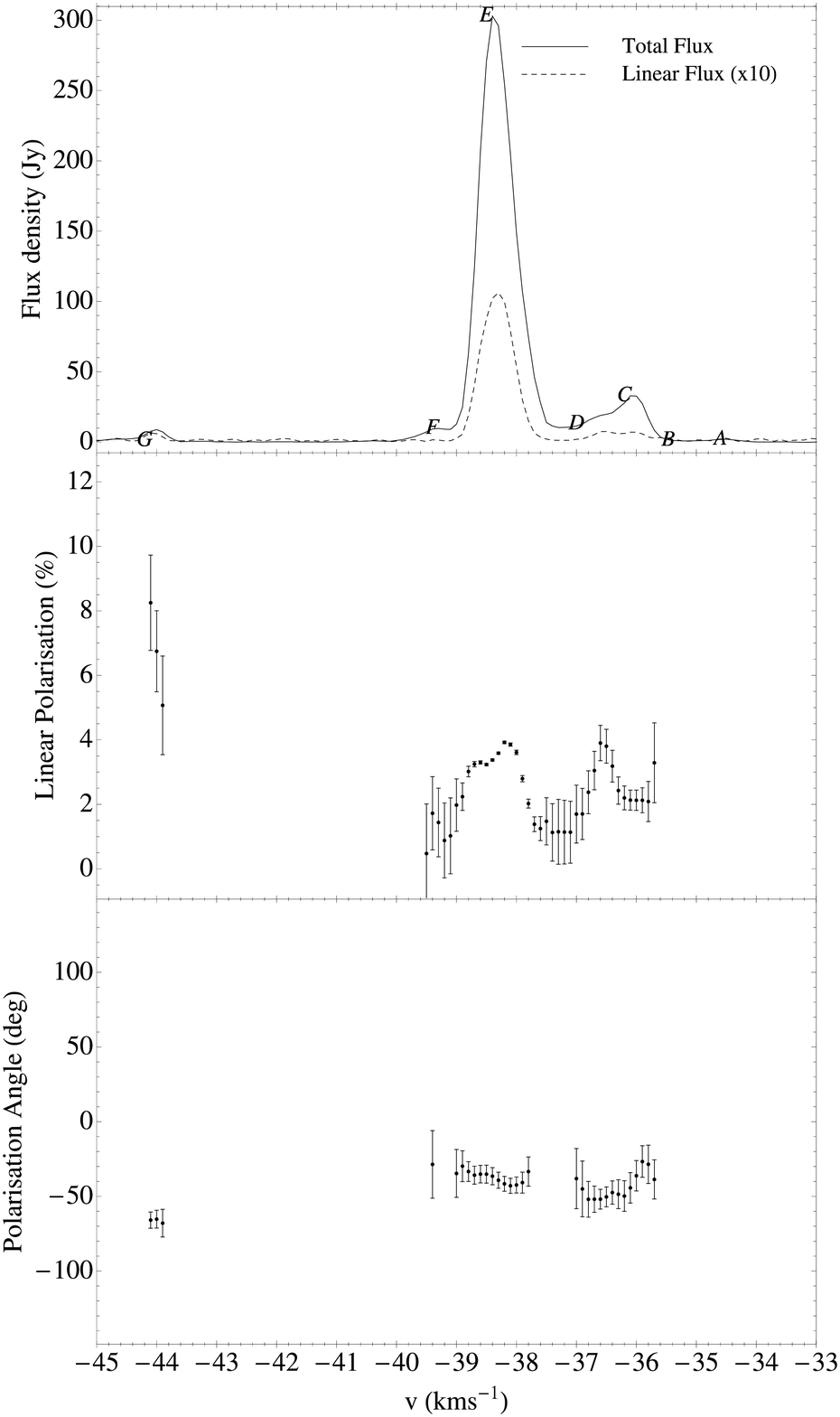}
\caption{As before, for G305.21$+$0.21, relative to 13:11:14.4 -62:34:26}\label{fig:1}

\end{figure}

The results from G305.21+0.21 are presented in Figure \ref{fig:1} in
the same form as for Figure \ref{fig:0}. 
This source displays relatively consistent polarisation angles across
the source, parallel to the linear distribution of the maser
features. This is inconsistent with the model of methanol masers
tracing an accretion disk and provides support for an outflow or shock
front model.  
The linear structure could be described as an ellipse, but the
polarisation angle neither shows a sweep with phase around the arc,
nor does the polarisation angle show a 90$^\circ$ phase change at the
centre. These would be expected if the source was a disk dominated by
infall or outflow.
The accretion disk model was supported for this source in a study by
\cite{Walsh06} who found that this source shares similar observational
properties with a low-mass star forming region in the accretion
phase. They suggested that this source may be in the very early stages
of massive star-formation, where the central mass is deeply embedded
in the molecular cloud.  Although they state that this source may
provide evidence that there is a short-lived accretion phase in this
early stage of massive star-formation, they were unable to find any
direct tracers of accretion processes \citep{Walsh06}.
\cite{deBuizer03} detected parallel H$_{2}$ emission from this source
that would appear to contradict the accretion disk model, instead
lending support to the outflow model. However, he also noted that
there is a second nearby methanol source, and it is confusing as to
which source can be associated with the H$_{2}$ emission. The parallel
major axis and magnetic field vectors found lend support to De
Buizer's association of the emission with an outflow.

\subsection*{G309.92+0.47}
\begin{figure}
\includegraphics[width=8cm]{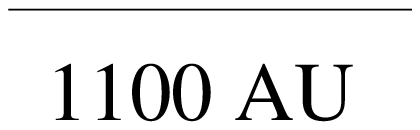}
%\end{figure}

%\begin{figure}
\includegraphics[width=8cm]{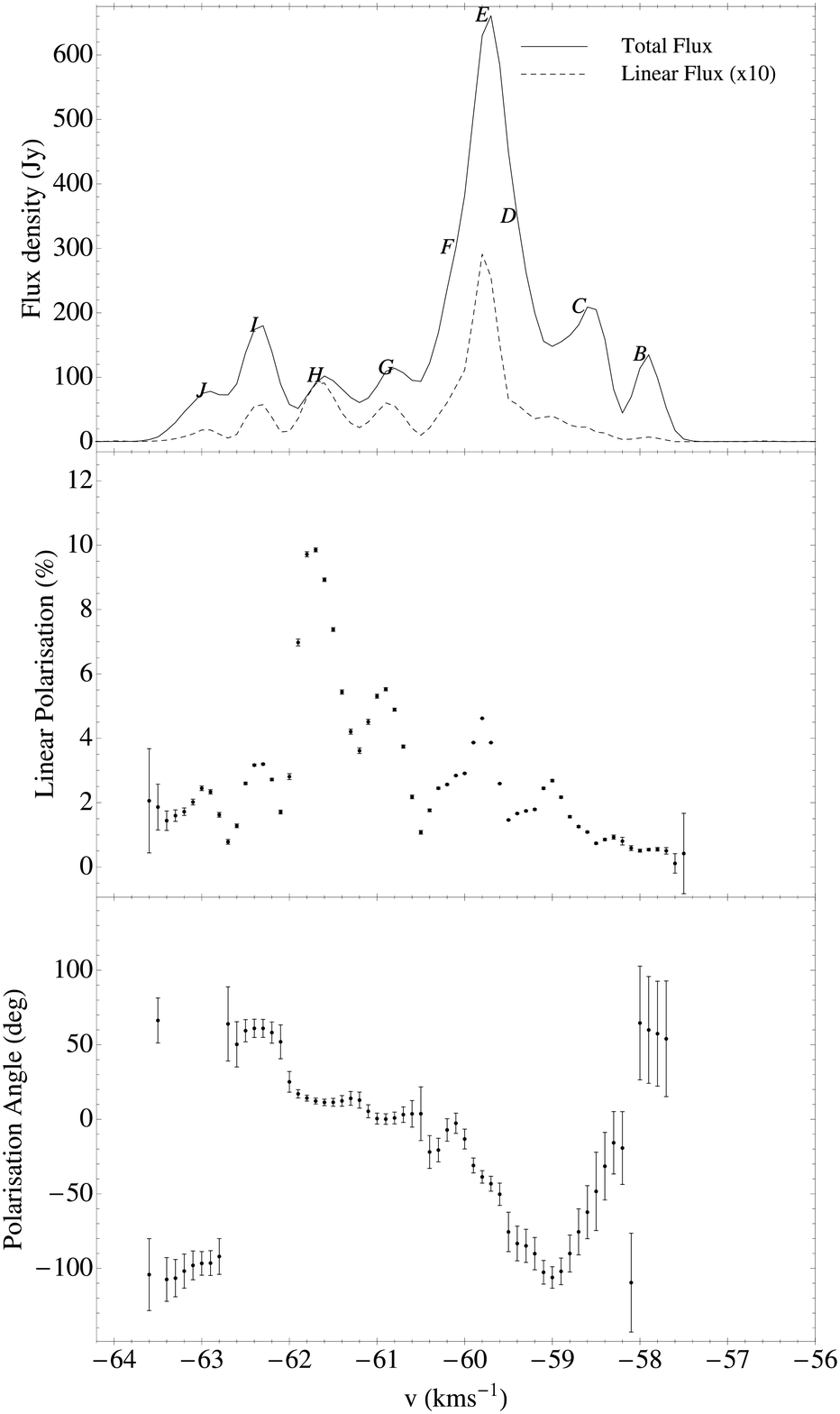}
\caption{As before, for G309.92$+$0.47, relative to 13:50:41.85 -61:35:11}\label{fig:2}

\end{figure}

The results from G309.92+0.47 are presented in Figure \ref{fig:2}, in
the same form as for Figure \ref{fig:0}.
The magnetic field morphology of this region is complex; vectors
appear to be predominantly perpendicular, but tend towards parallel in
the northern cluster (B-E) and feature J. Note that a weak feature A
appears at -54.2\,kms$^{-1}$ and thus has been clipped from the
spectra, but appears in the upper panel plot.
A morphology flip of polarisation was noted in a study by
\cite{Caswell95}, who identified Zeeman pairs in OH maser sources
G309.92+0.48 that indicated magnetic fields in opposite directions in
this associated region. This could be associated with components B--E
compared to the other features, but we note that the polarisation
angle changes smoothly across the source.
It is unclear whether the same field as for the OH would be
affecting the methanol masers, which may be offset from the OH
emission. 
The maser features themselves appear in an arc- or ring-like
distribution similar to those identified by \cite{Bartkiewicz09}.
however the polarisation angles do not show a jump such as which would
occur if the angle between the line of sight and magnetic field vector
passed through $\sim$55$^\circ$.
This source was originally considered to be a strong candidate for the disk
model, however \cite{deBuizer03} found no evidence of H$_{2}$ emission
perpendicular to the masers as would be expected in this
case. Although our field vectors support the disk model for some
features, generally it is not consistent. Our results also contradict
a shock or outflow model, where they would be expected to trace the
emission. It is possible they could lie in a region with variable
magnetic fields. 
%Alternatively, a new environment model may need to be
%found that accommodates magnetic variability across arc-like
%distributions.

\subsection*{G310.13+0.75}
\begin{figure}
\includegraphics[width=8cm]{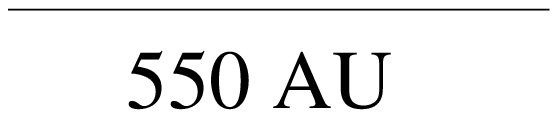}
%\end{figure}

%\begin{figure}
\includegraphics[width=8cm]{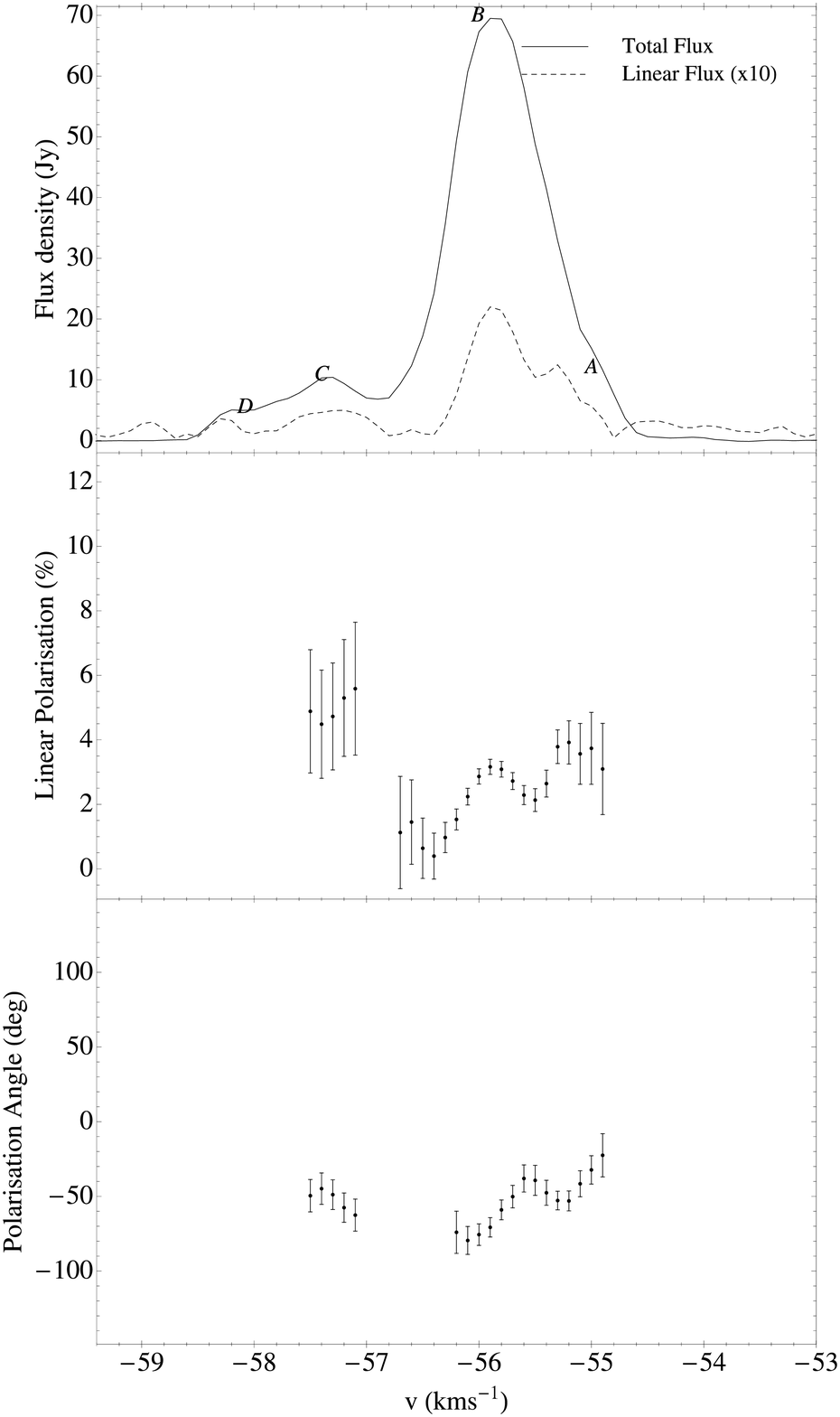}
\caption{As before, for G310.13$+$0.75, relative to 13:51:54.2 -61:16:18}\label{fig:3}

\end{figure}

The results from G310.13+0.75 are presented in Figure \ref{fig:3}, in the same
form as for Figure \ref{fig:0}.
G310.13+0.75 is another relatively weak source, peaking at 69.5
Jy. Whilst four velocity features are identifiable these do not
display a clear linear distribution, hence it is difficult to associate
a linear axis with this source. The polarisation angles found vary
across a 40 degree angle range, leading to a confused magnetic field
morphology for the source. No previous studies of the source were
available to consider additional properties of the region and this
source has been classified as possessing disordered structure. It is
possible that this could be a combined torus and outflow source. 

\subsection*{G316.64-0.08}
\begin{figure}
\includegraphics[width=8cm]{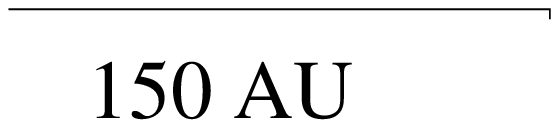}
%\end{figure}

%\begin{figure}
\includegraphics[width=8cm]{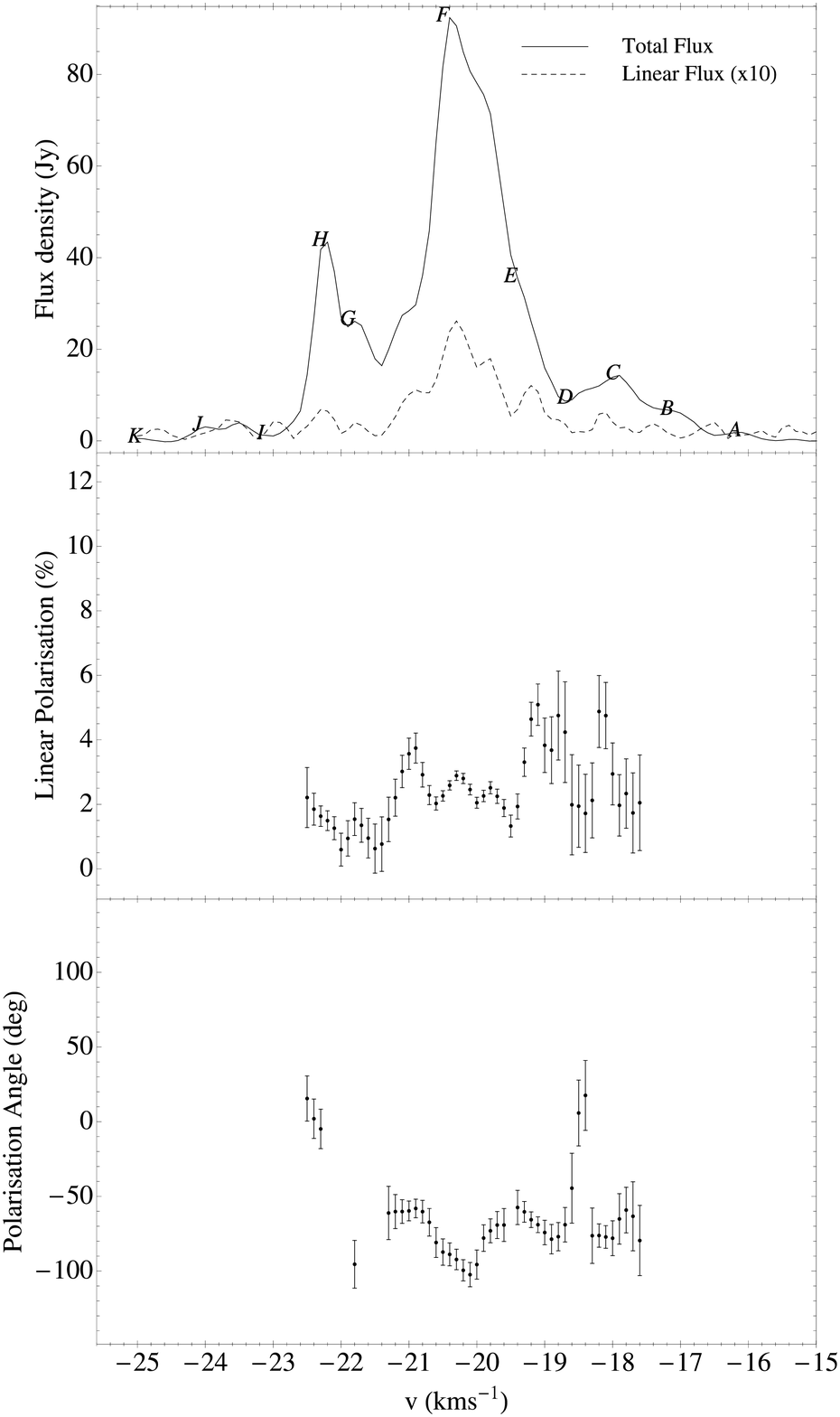}
\caption{As before, for G316.64$-$0.08, relative to 14:44:18.43 -59:55:12}\label{fig:4}

\end{figure}

The results from G316.64-0.08 are presented in Figure \ref{fig:4}, in the same
form as for Figure \ref{fig:0}.
The magnetic field vectors are aligned parallel to a linear axis as
ruled through components A through I. Two additional weaker components
J and K have been excluded from the position map, as they appear
offset from the primary cluster by $\sim$(-0.22,0.25) arcseconds and
may represent a separate emitting region, complicating the linear
fit. Component H shows an angle perpendicular to the magnetic field
trend, however the low fractional polarisation ($<$2\%) could be
affecting the accuracy of this angle measurement. Alternatively it
could be a region where the line of sight to magnetic field axis
crosses 55$^\circ$.  But the varying magnetic field vectors could also
indicate the maser clusters may not belong to the same environment and
it is possible that there may be more than one stellar mass in the
region. As the central cluster displays parallel morphology this may
indicate a further example of masers at a shock-front or outflow.

\subsection*{G335.79+0.17}
\begin{figure}
\includegraphics[width=8cm]{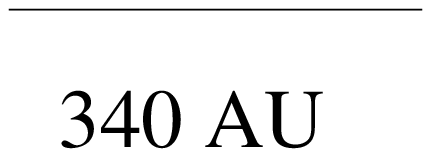}
%\end{figure}

%\begin{figure}
\includegraphics[width=8cm]{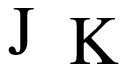}
\caption{As before, for G335.79$+$0.17, relative to 16:29:47.33 -48:15:52.4}\label{fig:5}

\end{figure}

The results from G335.79+0.17 are presented in Figure \ref{fig:5}, in the same
form as for Figure \ref{fig:0}.
This source contains a cluster of maser features in a linear
distribution, an outlying source consistent with an axis passing
through this cluster (H, F) and an offset region of weaker maser
emission (I, J, K).  The magnetic field vectors found for these
features emerge parallel to this axis, excluding component I in the
weaker cluster that emerges perpendicular to this trend.  A similar
distribution is seen in 12.2 GHz methanol maser emission at the same
site, and \cite{Phillips98} suggested the linear sources reflect a
disk, whereas the weaker features may be in a dense region of an
outflow. Although this is consistent with the field vector found for
feature I, it is inconsistent with the parallel morphology of the
other magnetic field angles. \cite{deBuizer03} detected H$_{2}$
emission parallel to this region, although it was unclear if this
could be directly associated with the methanol maser
emission. However, parallel H$_{2}$ emission is consistent with the
magnetic field angles if they in fact trace an outflow. The parallel
angles may also indicate consistency with the shock front model. The
weaker cluster complicates applying a shock front or outflow model
across the region, however these may represent a separate emitting
environment or secondary body.

\subsection*{G339.88-1.26}
\begin{figure}
\includegraphics[width=8cm]{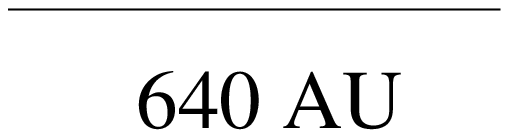}
%\end{figure}

%\begin{figure}
\includegraphics[width=8cm]{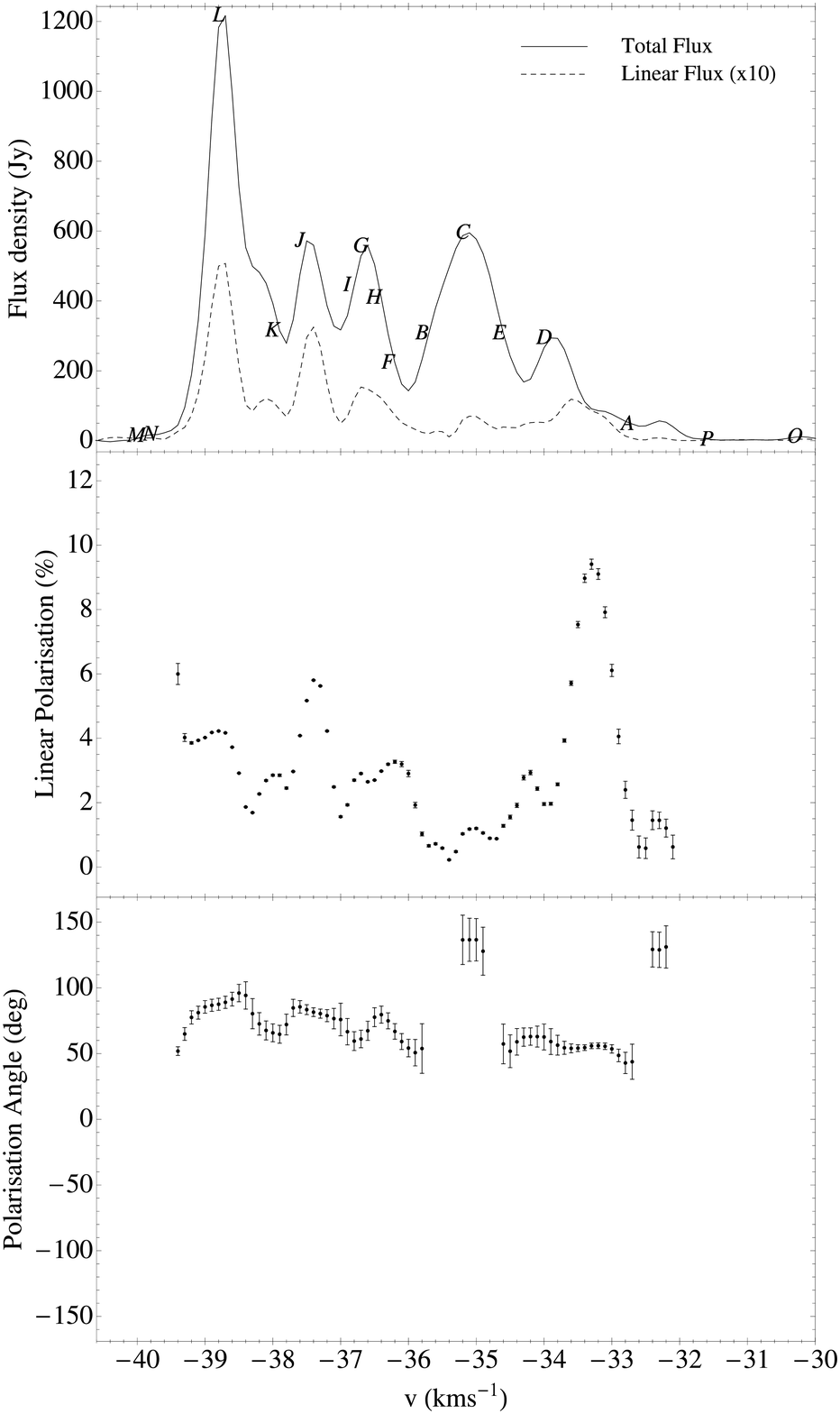}
\caption{As before, for G339.88$-$1.26, relative to 16:52:04.68 -46:08:34.4}\label{fig:6}

\end{figure}

The results from G339.88-1.26 are presented in Figure \ref{fig:6}, in the same
form as for Figure \ref{fig:0}.
This source was the subject of a thorough polarisation investigation
by \cite{Dodson08} using VLBI data that provided greater resolution
than is available in the datasets studied for this project. The
field vectors determined are consistent with those found previously
and appear to be predominantly parallel to the maser
emission. Features A and C display opposing polarisation vectors to
their neighbours, as observed in the VLBI results, which is most
likely from the change in the magnetic field line of sight passing through
55$^\circ$ to the observer, and flipping the magnetic field
orientation. The agreement of our averaged vectors with the overall
trends of the more detailed VLBI vectors supports the credibility of
our method as a test for large-scale structure. However, extended
observations at the resolution given by VLBI may provide a clearer
indication of finer magnetic field structure, which would be a
valuable probe of the inconclusive morphologies displayed across our
sources. This source has been classified as parallel and combined with
the previous results may be considered a candidate for the shock front
model.

\subsection*{G345.01+1.79}
\begin{figure}
\includegraphics[width=8cm]{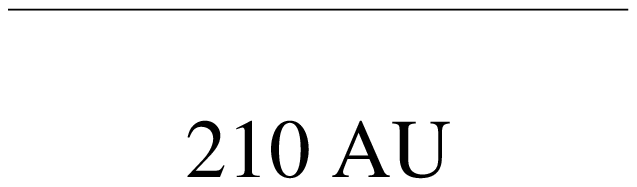}
%\end{figure}

%\begin{figure}
\includegraphics[width=8cm]{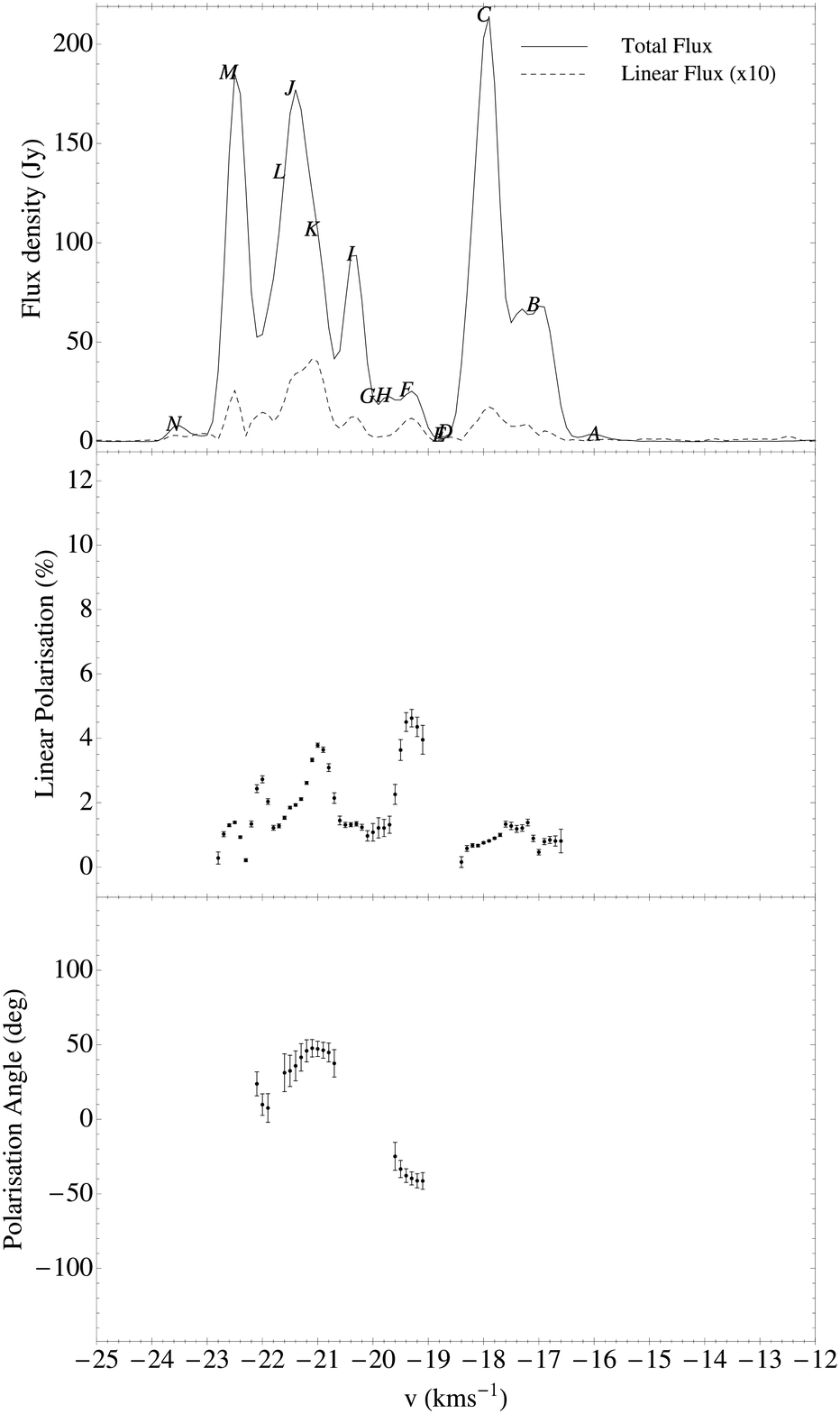}
\caption{As before, for G345.01$+$1.79, relative to 16:56:47.58 -40:14:25.9}\label{fig:7}

\end{figure}

The results from G345.01+1.79 are presented in Figure \ref{fig:7}, in the same
form as for Figure \ref{fig:0}.
G345.01+1.79 consists of maser features in a linear distribution, with
a clear velocity gradient along the source, as can also be seen in
\cite{Norris98}. The magnetic field morphology across the source is
unclear, with vectors F and H emerging parallel to the axis, whereas
vectors K, J and L are perpendicular.  
If this jump was due to the angle of the line of sight to the magnetic
field vectors the magnetic fields would have to be very disordered,
and switch from one domain to the other, twice. We have therefore
classified it as disordered in Table \ref{tab:Summary-of-Source}.
\cite{deBuizer03} identified extended infra-red emission offset from
the maser site but parallel to their distribution, as well as H$_{2}$
emission from the same elongated region. This is indicative of masers
distributed in an outflow. This is supported by the relatively high
collimation of the maser features, however the disordered magnetic
field vectors found are inconsistent with this model. The outflow
model also offers no clear explanation for the linear velocity
gradient of the source.  A further study of this source by
\cite{Salii03} states that there is extended evidence for shock fronts
propagating through this region, possibly due to the interaction of
stellar winds from the central mass with the surrounding molecular
cloud.
%The shock front model proposed
%by \cite{Dodson04} focuses on external shocks propagating inwards and
%it is unclear if this model may be more applicable to this region.
The central shocks suggested by Salii et al (2003) would require the young
protostar to have begun emission, which would indicate maser activity
at a later evolutionary phase than, for example, the early phase that
was suggested for G305.21+0.21 by Walsh and Burton (2006). Again, the
magnetic field vectors found do not show consistent support for this
model. It should be noted that many of the velocity features for this
source displayed low fractional polarisation ($<$2\%), hence it is
possible that the magnetic field across this source is weak and
disordered, complicating its association with the models being tested.

\subsection*{NGC6334F}
\begin{figure}
\includegraphics[width=8cm]{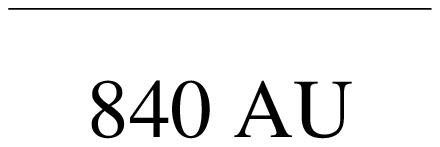}
%\end{figure}

%\begin{figure}
\includegraphics[width=8cm]{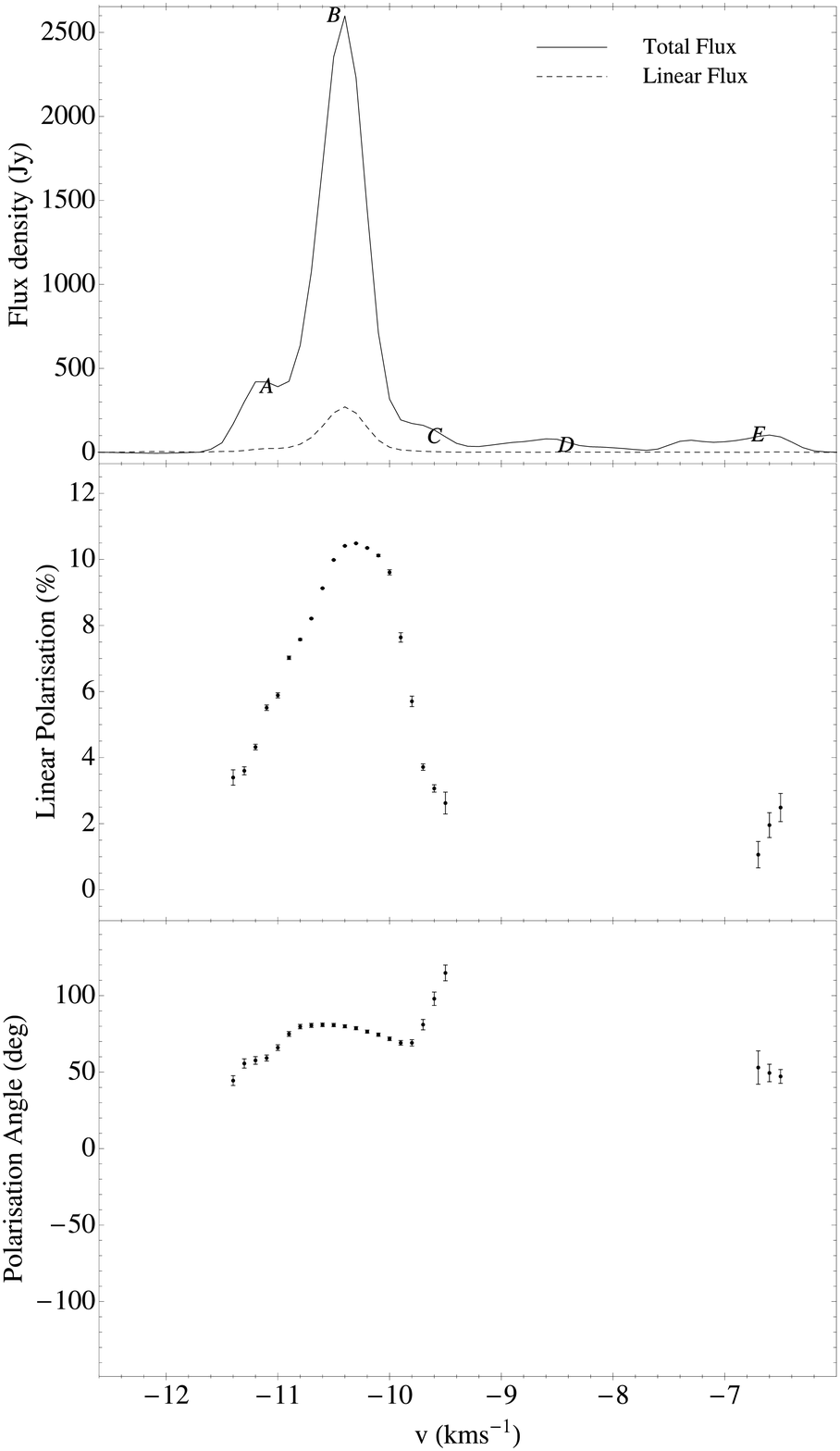}
\caption{As before, for NGC6334F (central), relative to 17:20:53.45 -35:47:00.5}\label{fig:8}

\end{figure}

The results from NGC6334F are presented in Figures \ref{fig:8} and
\ref{fig:9}, in the same form as for Figure \ref{fig:0}.
NGC6334F consists of two regions offset by 3.5 arcseconds.  We
found polarisation angles across the strongest maser features of each
source that agree with those found by  \cite{Ellingsen02}.
% We did not have 
Within the central region, the magnetic field emerges parallel to the
emission, whereas in the NW region the field vectors appear
perpendicular to the maser features. A weak feature, N, appears offset
to the north of the NW cluster, outside the range of the position
map. The entire NGC6334 region is a site of intense star formation
activity, with estimates suggesting the site houses 20-30,000 young
stars, dozens of which are presumed to be high-mass. The NGC6334F
region is associated with an ultra-compact H$_{2}$ region believed to
be due to a single massive protostar \citep{Feigelson09}. It
is, however, possible that the offset region represents emission
around a second system, considering the nature of the NGC6334
complex. Indeed this region of the complex has no signs of 7-mm
emission \citep{Carral97} indicating that it is significantly younger. 
The varying morphologies seen could be indicating that the
masers are emitting from different local environments. Alternatively,
if feature E is considered to be a separate region from A-C, these
vectors could be considered perpendicular to their emission suggesting
the central and NW regions may both be evidence of masers tracing
accretion discs around different stellar masses.

\section[Discussion]{Discussion}

\subsection{Disk with Outflow or Infall}

\cite{Bartkiewicz09} strongly suggest that methanol masers originate
in the disk or torus around a proto- or a young massive star where the
kinematics are strongly influenced by outflow or infall.
They base these deductions on their discovery of richer structures
(arcs and rings) that they uncovered due to the improved sensitivity
of their EVN observations.
%
%They warn in their paper that there is no straighforward explanation of
%the origin of the emission seen. %, and we must agree with them. 
The model of disk dominated by infall/outflow would produce a very
clear signature in the polarisation angles.
If the masers trace a disk dominated by infall or outflow the magnetic
field would be expected to lie in the disc, radially from the centre
of mass.
If the line of sight to the observer was from above, then the viewing
angle would be close to 90$^\circ$ and the magnetic field vectors
would be perpendicular to the observed linear polarisation vectors.
If the viewing angle was close to edge-on then the angle between the
fields and the line of sight would be greater than 55$^\circ$ at the
outer edges of the disk and less than 55$^\circ$ close to the centre
of mass.
The observable outcome would be either in the rotation of the
polarisation vectors around the centre if observed from above or, if
in an edge-on view, the fields would appear perpendicular to the
source structure at the extremities with a 90$^\circ$ flip approaching
the centre of mass.  We find no indications of this kind of structure
in the sources we have observed.

\subsection{Rotating Accretion Disk model}

If the masers trace a rotating accretion disk, which is observed edge on, the
magnetic fields would be expected to be perpendicular to the source
structure. As the line of sight would be close to 90$^\circ$ the magnetic
field vectors would be perpendicular to the polarisation vectors.

Whilst earlier polarisation studies found field vectors that
contradicted the accretion disk model, three of our sources have
fields that suggest that this model may still be applicable
(G291.27-0.70, G309.92+0.47, NGC6334F (NW)). It is possible that, as
suggested by \cite{Walsh06}, accretion disks may be present in the
early stages of high-mass star formation but may cease and give way to
other processes, for example after radiation pressure from the central
star begins to counteract the gravitational forces influencing
accretion. This suggestion is compatible to an extent with a recent
simulation by \cite{Krumholz09}. They found accretion disks to be
present during the early stages of high-mass star formation, but over
time these formed secondary stellar bodies that broke up the original
disks. 

\subsection{Shock Front Model}

If the masers trace shock fronts the magnetic fields would be expected
to be parallel to the shock front as the magnetic fields become `piled
up' along that front. 
Shock fronts tend to be observed close to edge-on where there is a
significant optical depth (e.g. \cite{hester87}), therefore the angle of the
field to the line of sight would be close to 90$^\circ$ and the magnetic
field vectors would be perpendicular to the polarisation vectors.
The parallel fields observed in five of our maser sources are in
agreement with such models, but the other five are not. 

In \cite{Dodson04} we suggested externally generated shocks
propagating through a rotating cloud could lead to the linear maser
distributions and velocity gradients that have been observed. We
postulated as a test of this theory that magnetic fields are expected
to be found parallel to the major axis of the masers, the genesis of
this paper.  A similar theory was suggested by \cite{Salii03} in a
discussion of the source G345.01+1.79, in which shocks were instead
occurring at the interface between stellar winds from the central mass
and the surrounding molecular cloud.
However only half of the sources observed show the predicted structure. 

\subsection{Outflow models}

If the masers trace outflows the 
%
%The 
magnetic fields would align with the material flow, tracing the
outflow direction.  When these are observed from the side the angle of
the field to the line of sight would be close to 90$^\circ$ and the
magnetic field vectors would be perpendicular to the polarisation
vectors. Alternatively when the outflow is directed towards or away
from the observer the angle is less than 55$^\circ$ and the vectors
would be parallel.

That there is no dominant field orientation in our maser sources could be
compatible with the model of masers located in outflows, as
outflows might be expected to have random alignment.
On the other hand we would expect outflows pointing close to our line
of sight to have a higher differential velocity across the
source. There is no obvious relationship between $\Delta$velocity and
angle $\theta_B$-$\theta_{axis}$ as seen in the data presented in
Table \ref{tab:Summary-of-Source}.

It is becoming possible to include outflows in sophisticated
simulations for Massive Star Formation, as investigated in
\cite{Cunn11}.  These simulations point to a reduction in the
radiation pressure on the infalling matter, potentially explaining why
Massive Stars can form despite the nominal Eddington limitations. At
the same time they predict winds of the order of 100 km\,s$^{-1}$ that
are much greater than those seen across the methanol maser sources. Of
course there are regions in the simulations where the velocities are
much lower and detailed analysis of where the masers may arise would
resolve these questions.

\subsection{Whither now?} %%% Better title %%%% Possible/Future directions for investigations/research

As discussed, we observed both perpendicular and parallel morphologies
between magnetic field vectors and maser distributions in our sample
of sources studied.  Although in several cases magnetic field vectors
agree with the postulation that methanol masers are tracing accretion
disks, the majority of observations cannot be explained by this model.
On the other hand the shock models would produce magnetic fields
perpendicular to the source structure, but this model has no
explanation for the parallel alignments.
Outflows could be aligned randomly to the line of sight. They would
have magnetic fields perpendicular to the source structure, but the
linear polarisation may or may not be perpendicular to the magnetic
field, preventing the observation of a single dominant tendency. This
is what we have discovered, but this model is not supported by the
expectation that the sources aligned parallel would have higher
velocities across the source.
If masers arise on a disk dominated by infall or outflow, and we
observed this from above, we would expect to measure a smooth sweep of
polarisation angle around the centre of the source. This is not
found. Nor do we see a reversal of the polarisation vectors
approaching the centre. This would be expected from such disks
observed edge on, due to the change in angle of the line-of-sight and
the magnetic fields.

It is possible that this lack of a dominant simple model may be an
indication that methanol masers are seen across a range of different
evolutionary stages. Alternatively, masers could be forming in
environments that cannot be described by simplistic general models,
such as the suggestion in \cite{Surcis11} that the masers form at the
interface between a combination of the two models: a torus and an outflow. 
The former is complicated to test, but the latter we believe can be
tackled by simulations.

This is timely as simulations are approaching a useful degree of
sophistication. Starting with \cite{Banerjee07} and then
\cite{Krumholz09} adaptive mesh refinement 3D
radiative-hydrodynamic simulations have shown the production of both
sustained accretion disks and multiple, lower mass, bodies in a
complex interplay as the MSF regions developed. Our simplistic models
for magnetic field alignments would not have produced an adequate
description of these sources. However these are only simulations and
included neither outflows nor magnetic field effects.  Outflows have
been included the work of \cite{Cunn11}. Magnetic field effects are
included in lower mass stellar formation simulation studies that use
Smoothed Particle Hydrodynamics \citep{Price09}.

None of these datasets can yet be used directly to identify the
regions that would produce maser emission, but approaches exist which
can be added to the existing simulations.
We propose to analyse new MSF simulations which we are performing, to
identify those regions where the pumping occurs and the masing
conditions are met.  We will be able to compare these predictions to
the structures actually observed, and their development with
evolution. We will be able to derive (with those simulations which
include the magnetic fields) the observed polarisation properties and
test the simulations against the observed reality.
The combination of simulations of these complex regions and detailed
observations of actual methanol masers offers a solution to the
longstanding lack of clarity in the interpretation of methanol maser
environments.

In addition new observational data, with reliable Stokes V data, will
also be required. Stokes V will allow direct estimates of the emerging
brightness temperatures, among other parameters. The improved C-band
receivers currently being installed at ATCA will greatly assist in
providing this. Furthermore higher resolution observations will allow
measurements of the spot sizes and provide the separation of
separate components at the same velocities.

\section{Conclusions}
We have presented an analysis of the polarisation properties of ten
6.67-GHz methanol maser sources observed by the ATCA, mapping the
magnetic field against the distribution of maser features. This
represents a large increase in the number of methanol maser sources
mapped in polarisation at high resolution.  We have utilised the
observed angles as a test of the theory that linear distributions of
methanol masers trace accretion disks in high-mass star forming
regions. Whilst three of the sources studied displayed magnetic fields
perpendicular to the major axis, as expected from an accretion disk,
five sources displayed parallel vectors more consistent with shock
fronts or outflows. 
None of the fields were found to be consistent with disks dominated by
infall or outflow.
Whilst outflows could produce both perpendicular and parallel
alignments, they should also show greater velocity ranges for the
latter case, which is not seen.

The sources observed to have magnetic fields parallel to maser
distribution agree with previous polarisation studies that similarly
found parallel morphologies, which cast doubt on the accretion disk
model \citep{Vlemmings06,Dodson08,Surcis09}.  The appearance of
perpendicular vectors in our results may be an indication that
methanol masers are being emitted from different environments
or different stages of stellar evolution, wherever favourable
conditions for maser action occur. 
Further work will be undertaken to examine whether simulations will be
able to explain the observed features. 

%\onecolumn

\section*{Acknowledgments}

Dr Phillips provided the majority of the maser spot positions from
data published in \cite{Phillips98}. 
Dr Brown and Dr Van Eck provided the solutions for the RM from their model
of the Galactic Magnetic field, for each source direction and
distance.
We are grateful for the comments of the anonymous referee were
extremely helpful.
The 2008 observations were made under Directors Time at the Australia Telescope
for this project. Further data was extracted from the Australia
Telescope Online Archive. The Australia Telescope is funded by the
Commonwealth of Australia for operation as a National Facility managed
by CSIRO.

%\bibliographystyle{mn2e}
%\bibliography{moriartyrefs}

\twocolumn
%\clearpage
%\begin{minipage}[t]{0.5\textwidth}
%\end{minipage}
\hfill
\begin{minipage}[t]{0.45\textwidth}
%% %\begin{figure}
%% %\caption{
\vspace{2cm}
\noindent {Top) The map of the maser emission for G291.27$-$0.70,
  relative to 11:11:53.37 -61:18:23.50 with the integrated features
  labeled with letters and, where polarized emission was detected,
  overlaid with the magnetic field direction scaled by the linear
  polarized flux. One $\sigma$ errors are indicated and the lower
  horizontal bar provides the scale for 1\% linear polarisation. The
  spot colour indicates the velocity and the light solid line
  indicates the best fit axis to the maser features. The upper
  horizontal bar provides a scale in AU, using the distances we have
  adopted and listed in Table \ref{tab:Summary-of-Source}.
   Bottom) The velocity spectra of the masers, with the integrated
   features labeled with letters. The top panel is the total flux
   density and the linear polarized flux density, scaled-up by a factor
   of 10. The measured fractional linear polarisation, and one sigma errors are
   shown in the middle panel. The derived polarisation angle $\chi$, with one
   sigma errors, is shown in the bottom panel. }
%% %}\label{fig:0}
%% %\end{figure}
\end{minipage}

%\twocolumn

%\clearpage

\clearpage

\onecolumn
\begin{table}
\begin{tabular}{ll|rrrr|rrrrrr}
Name   & Classification & Flux &  Axis & 
                       Av. pol. & V$_{lsr}$ & Distance & RM & %Rotation  & 
Offset & Linear & $\Delta$ velocity\\ 
& & & $\theta_{axis}$ & angle $\chi$ & & &  & %angle $\phi$ & 
 $\theta_B - \theta_{\rm axis}$ & Size& \\
& & (Jy\,km\,s$^{-1}$) & (deg) & (deg) & (km\,s$^{-1}$) & (kpc) & (rad\,m$^{-2}$) & %angle $\phi$ & 
 (deg) & (AU) & (km\,s$^{-1}$)\\
\hline
%%\startdata
% \tableline
G291.27$-$0.70 & perpendicular & 40 &-77&-32$\pm$5 &-29.5 &3.1&51.1&%5.9 & 
-51 & 0.6 & 2.9 \\%100*3.1&2\\
G305.21$+$0.21 & parallel & 190 &48& -51$\pm$14&-38&4.0&99.9&%11.6 &
-20 & 1.7 & 11.5\\%300*4&10\\
G309.92$+$0.47 & perpendicular & 898 &35& 2$\pm$56&-60&5.5&-112.6&%-13.1 &
70 & 4.7 & 4.9\\%800*5.5&6\\
G310.13$+$0.75 & disordered & 81 &-&-56$\pm$20 &-56 &5.5&-113.9&%-13.2 &
 - & 0.6 & 3.8\\%100*5.5&4\\
G316.64$-$0.08 & parallel & 102 &34 &-67 $\pm$36&-20.5&1.5& 15.8&%1.8 & 
-13 & 0.8 & 8.8 \\%400*1.5&6\\
G335.79$+$0.17 & parallel & 252 &-69 &44$\pm$28&-48&3.4&-38.4&%-4.5 & 
28 & 1.3 & 10.6 \\%600*3.4&10\\
G339.88$-$1.26 & parallel & 2571 &-60 &77$\pm$24 &-39 &3.2&-18.3&%-2.1 & 
48 & 4.3 & 9.7 \\%1000*3.2&9\\
G345.01$+$1.79 & disordered & 396 &74 &5$\pm$39 &-18 &2.1& 8.4&%1.0 & 
20 & 0.8 & 7.6\\%300*2.1&8\\
NGC6334F (central) & parallel & 1587& -41&77$\pm$20 &-10.5 &2.1&-0.4&%-0.0 & 
28 & 4.7 & 4.4\\%1000*2.1&6\\
NGC6334F (NW) & perpendicular &733 &-80 &-71$\pm$20 &-11 &2.1&-0.4&%-0.0 & 
-81 & 4.6 & 5.0 \\%600*2.1&3\\
%\enddata%\tableline
\hline
\end{tabular}
%\end{deluxetable}
%\end{center}
%\end{minipage}
   \caption{{\bf Parameters of the Methanol Masers}: Column 1 gives the
     source name. Column 2 our classification based on the orientation
     of the magnetic field directions of the individual maser
     components (labeled with letters in the figures) with respect to
     the major axis of the source. Column 3 gives the integrated flux
     density (in Jy\,kms$^{-1}$) across the source. Column 4 gives the
     fitted major axis to the individual maser components
     positions. Column 5 gives the averaged polarisation angle (weighed
     by the errors) for the individual masing components, with the
     standard deviation around that value. Column 6 gives the local standard of
     rest velocity of the peak emission of the maser. Column 7 gives the distance
     used to calculate the RM.
     Column 8 gives the RM derived from the
     model of Van Eck and Brown (2010), which is scaled by 0.116 to convert to
     $\phi$ in degrees of rotation at 6.67\,GHz.
     Column 9 gives the offset between the major axis and
     rotated averaged magnetic field direction. Column 10 gives the linear size, derived from the angular extent of the observed emission scaled with the distance. Column 11 gives the range of velocities seen in across the source emission.}
\label{tab:Summary-of-Source}
\end{table}

%\label{lastpage}

\end{document}